\documentclass[pra,aps,groupedaddress,twocolumn,floatfix,nofootinbib]{revtex4-2}
\usepackage{graphicx}
\usepackage{soul}
\usepackage{float}
\usepackage{amsthm, amsmath, amsfonts}
 \usepackage[normalem]{ulem}

\newcommand{\beq}{\begin{equation}}
\newcommand{\eeq}{\end{equation}}
\newcommand{\beqa}{\begin{eqnarray}}
\newcommand{\eeqa}{\end{eqnarray}}

\newcommand{\<}[1]{\left\langle #1 \right\rangle}

\def\half{\frac{1}{2}}
\def\opone{\leavevmode\hbox{\small1\normalsize\kern-.33em1}}

\usepackage{xcolor}
\definecolor{lasallegreen}{rgb}{0.03, 0.47, 0.19}

\begin{document}

\title{Non-Local Boxes for Networks}
\author{Jean-Daniel Bancal$^{1,2}$ and Nicolas Gisin$^{1,3}$ \\
\it \small   $^1$Group of Applied Physics, University of Geneva, 1211 Geneva 4,    Switzerland \\
$^2$Université Paris-Saclay, CEA, CNRS, Institut de physique théorique, 91191, Gif-sur-Yvette, France\\
$^3$Schaffhausen Institute of Technology - SIT, Geneva, Switzerland}

\date{\small \today}
\begin{abstract}
Nonlocal boxes are conceptual tools that capture the essence of the phenomenon of quantum non-locality, central to modern quantum theory and quantum technologies. We introduce network nonlocal boxes tailored for quantum networks under the natural assumption that these networks connect independent sources and do not allow signaling. Hence, these boxes satisfy the No-Signaling and Independence (NSI) principle. For the case of boxes without inputs, connecting pairs of bipartite sources and producing binary outputs, we prove that the sources and boxes producing local random outputs and maximal 2-box correlations, i.e. $E_2=\sqrt{2}-1$, $E_2^o=1$, are essentially unique.
\end{abstract}
\maketitle

\section{Introduction}\label{intro}
Non-locality is a key feature of quantum physics and one of the major discovery - arguably the major discovery - of last century physics. Modern quantum technology promises, in addition to quantum computers, quantum networks that will connect these quantum processors and offer proven confidentiality of all communications. It is thus natural and timely to study quantum non-locality in networks, a field that has been burgeoning for about a decade under the names of bilocality and $n$-locality, for 2 and $n$ independent sources, respectively \cite{Branciard, Fritz,Branciard2,ChavesFritz,Tavakoli,Henson,Wood,ChavesKueng,TavakoliConnected,Fritz2,Rosset,
Chaves,Tavakoli3,Tavakoli2,Andreoli,Fraser,Luo,Inflation,Wolfe,Salman,Renou}. Let us stress that in networks the characteristic feature of quantum physics, namely entanglement, enters twice. First, entanglement of the subsystems emitted by the sources. Second, entanglement produced by the joint measurements that connect independent subsystems emitted by different sources. This second form of entanglement is much less studied and understood than the first one \cite{EJM}.

In this work we go beyond quantum physics and study $n$-locality for arbitrary boxes only limited by no-signaling and independence, a principle we name NSI \cite{GisinBancal}. This is in the spirit of the non-local boxes introduced by Popescu and Rochlich - the so-called PR-boxes \cite{PR} - as a conceptual tool to study Bell non-locality. Here, however, we go beyond Bell non-locality, our boxes connect independent sources and have no inputs. A priori, such network non-local boxes could have any number of outputs and connect any number of sources, with sources connected to any number of boxes. In this paper we restrict ourselves to boxes connecting two sources, producing binary outputs and sources connected to two boxes, see Fig 1. We name the networks obtained with these resources binary networks. Similarly to the PR-boxes the aim is the set the limits of non-locality in networks imposed by the NSI principle and to offer conceptual tools to study this new form of non-locality.

Assuming that locally each output of a box is totally random and that the outputs of two neighboring boxes are maximally correlated when using a particular type of source, we find that the statistics produced by this box in presence of this source in a binary network are unique (up to flipping all outputs).

\begin{figure}[h]
\centering
\includegraphics[width=1\columnwidth]{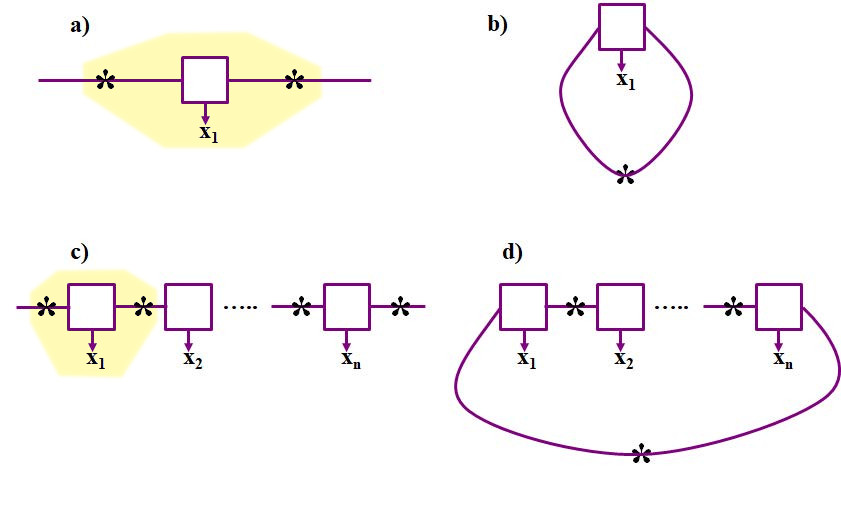}
\caption{\it Various networks. Boxes are indicated with a square that outputs $x_j$. Sources are marks as {\Large$*$}. A priori, the sources and boxes could be asymmetrical (i.e. non invariant under exchange of their arms. Since this turns out to be irrelevant here, we simply depict them in a symmetric fashion.}
a) One box in a line. b) One box in a loop. c) $n$ boxes in a line. d) $n$ boxes in a loop.\\
The NSI principle states that the statistics produced in situations which are indistinguishable to the parties must be identical, such as the statistics of $x_1$ in situations a) and c), by identification of the yellow regions (see~\cite{GisinBancal} for more details). Note that this principle does not impose the statistics of $x_1$ to be identical in b), however, because the change from two sources in a) to just one source in b) can be detected by the party when producing $x_1$, i.e. no equivalent of the yellow region in a) can be found in b).
\end{figure}

\section{Binary Network Non-Local Boxes}
First, consider two identical but independent sources, sending out subsystems in two opposite directions, coupled by one box as illustrated in Fig. 1.a. Each box ouputs a bit $x_j=\pm1$, which we often label merely $\pm$. We assume that this bit is random, hence its expectation value is zero: $E_1=\<{x_j}=0$. Figure 1.b illustrates the case of a box connected to the two parts of a single source. Here also we assume $E_1^o=\<{x_j}=0$, where the superscript ``o" indicates that this correlator corresponds to a closed loop.

Next, let us add more independent sources and boxes. We focus on the case in which all boxes are identical, i.e. copies of our box of interest, and similarly all sources are also copies of our source of interest. Only two fully connected configurations are possible. Either all sources and boxes are on a line, Fig. 1c, or they form a loop, Fig. 1d. It is important to realize that no other configurations is possible. Hence, proving the existence of a source and box compatible with these two configurations suffices to prove that they are conceptually possible.

In both configurations, the probability of the outputs $x_j$ can conveniently be expressed in terms of all correlators. Since all sources and boxes are identical, the correlators between $k$ adjacent boxes are all equal, denoted $E_k=\<{x_{m+1}\cdot x_{m+2}\cdot ...\cdot x_{m+k}}$. In case of a polygon configuration with $n$ boxes at the vertices, i.e. a loop, there is one additional correlator: $E_n^o=\<{x_1\cdot x_2\cdot...\cdot x_n}$.

Let us consider some examples. First, the joint probability of outputs for 3 boxes on a line reads:
\beqa
p_3(x_1,x_2,x_3)&=&\frac{1}{2^3}\big(1+(x_1+x_2+x_3)E1 \nonumber\\
&+& (x_1x_2+x_2x_3)E_2 \nonumber\\
&+& x_1x_2x_3E_3 + x_1x_3E_1^2\big) \label{p3a}
\eeqa
where the bipartite correlator associated to $x_1x_3$ is $E_1^2$ due to the independence assumption. Since by assumption $E_1=0$, we obtain
\beqa
p_3(x_1,x_2,x_3)&=&\frac{1}{2^3}\big(1+ (x_1x_2+x_2x_3)E_2 \nonumber\\
&+& x_1x_2x_3E_3\big) \label{p3}
\eeqa
where the first term guarantees normalization. Next, for 4 boxes in a square we have:
\beqa
p_4^o(x_1x_2x_3x_4)&=&\frac{1}{2^4}\big(1+(x_1x_2+x_2x_3+x_3x_4+x_4x_1)E_2 \nonumber\\
&+&(x_1x_2x_3+x_2x_3x_4+x_3x_4x_1+x_4x_1x_2)E_3 \nonumber\\
&+& x_1x_2x_3x_4E_4^o\big)
\eeqa
and so on. 

The correlators corresponding to disconnected sets factorize, because of the assumed independence of the sources, as in the last term in (\ref{p3a}). It is important to realize that for hexagons and 5 boxes in a line and larger networks, independence implies some non-linear correlators. For example, in a line with 5 boxes the last correlator equals the square of the 2-box correlator:
\beqa
p_5(x_1...x_5)&=&\frac{1}{2^5}\big(1+\sum_{k=1}^4x_kx_{k+1}E_2 +\sum_{k=1}^3 x_kx_{k+1}x_{k+2}E_3 \nonumber\\
&+& \sum_{k=1}^2 x_kx_{k+1}x_{k+2}x_{k+3}E_4 + \prod_{k=1}^5 x_k E_5 \nonumber\\
&+& x_1x_2x_4x_5 E_2^2\big)
\eeqa

Obviously, the correlators have to be such that $p(x_j)\geq0$ for all outputs strings $\{x_j\}$. In \cite{GisinBancal}, using 6 sources and 6 boxes in an hexagonal loop, we proved the following bound on the 2-box correlator:
\beq\label{sqrt2m1}
E_2\leq\sqrt{2}-1\approx 0.4142
\eeq
Here we prove that this upper bound can be saturated for all possible binary networks  and that the saturation of (\ref{sqrt2m1}) is achieved in an essentially unique way (up to flipping all outputs). 

Our main results are the correlators for boxes in a line (\ref{En}), section \ref{line}, and in a loop (\ref{Eno}), section \ref{loop}. These correlators are unique given $E_1=0$, $E_2=\sqrt{2}-1$ and $E_2^o=1$, and we prove that they guarantee that all probabilities $p(\vec x)$ and $p^o(\vec x)$, where $\vec x=(x_1..x_n)$, are non-negative. But first, in the next section, we consider 3 boxes in a line and prove that the value of the corresponding correlator $E_3$ is unique (up to its sign).

\section{3-box correlator}\label{E3}
Consider 3 boxes in a line as in eq. (\ref{p3}). We have $p(+-+)=\frac{1}{8}\big(1-2E_2-E_3\big)\geq0$. Hence, $E_2=\sqrt{2}-1$ imposes $E_3\leq1-E_2=3-2\sqrt{2}$. Interestingly, when considering lines with more copies of the box, this inequality becomes an equality (up to a sign flip), namely
\beq\label{E3}
E_3=\pm(3-2\sqrt{2}),
\eeq
as implied by the following lemma proven in the Appendix.

\vspace{0.3cm}\!\!\!\!\!\!\textbf{Lemma 1.}\quad \textit{In a line configuration with 7 boxes, whenever $E_1=0$ and $E_2=\sqrt{2}-1$, the following inequality holds:}
\begin{equation}\label{E3bound}
E_3^2 \geq (3-2\sqrt{2})^2.
\end{equation}
\vspace{0.2cm}

Henceforth we assume the positive sign for $E_3$. We'll see that this choice implies that all correlators are non-negative. Note that by flipping all outputs, all odd correlators, in particular $E_3$, merely change sign.

The relation (\ref{E3}), inserted in (\ref{p3}), has the following important consequence: $p(+-+)=0$. Accordingly, there are never 3 adjacent boxes with outputs $+-+$:
\beqa
p_n(x_1..x_{j-1}+-+x_{j+3}..x_n)&=& \nonumber\\
p(+-+)&\cdot& \\
p(x_1..x_{x-1}x_{j+3}..x_n|x_j=+,x_{j+1}&=&-,x_{j+2}=+)=0 \nonumber
\eeqa
This in turn implies that whenever two connected boxes output $+-$, then the next box necessarily outputs $-$. This is key in proving existence and uniqueness of our network non-local box, as we show in the two next sections.

\section{Recursion formula for the $E_k$}\label{line}
Let us define $q_n(\vec x)=2^np_n(\vec x)$, so that the normalization factors drop out. We still have the non-negativity conditions $q(\vec x)\geq0$ for all $\vec x$. The case of $n+1$ boxes in a line can be reduced to $n$ boxes as follows:
\beqa
q_{n+1}(x_1..x_{n+1})&=&q_n(x_1..x_n) + (\Pi_{j=1}^{n+1} x_j) E_{n+1} \nonumber\\
&+& \sum_{k=0}^{n-1}(\Pi_{j=k+2}^{n+1}x_j)q_kE_{n-k} \\
&=&q(x_1..x_n) \nonumber\\
&+& (\Pi_{j=1}^{n+1} x_j) (E_{n+1}+Q_n) \label{qnp1}
\eeqa
where $Q_n=\sum_{k=0}^{n-1}(\Pi_{j=1}^{k+1}x_j)q_kE_{n-k}$, $q_k=q_k(x_1..x_k)$ and we used $x_j^2=1$ for all $j$. The first term on the left of (\ref{qnp1}) takes into account the first $n$ boxes, the second term all $n+1$ boxes and the last term all cases that involve the last box but not all boxes. 

Considering successively the cases $(\Pi_{j=1}^{n+1} x_j) = \pm1$ one gets:
\beq \label{EnBound}
q_n(x_1..x_n)-Q_n\geq E_{n+1}\geq -q_n(x_1..x_n)-Q_n
\eeq
Apply this to the string of alternating outputs $q_n(+-+-+-...)$, the result of the previous section implies $q_k=0$ for all $k\geq3$. Hence, from (\ref{qnp1}) and (\ref{EnBound}) one obtains the recursion formula:
\beq\label{Enp1}
E_{n+1} = -E_n + E_{n-1} + (1-E_2)E_{n-2}
\eeq

This leads to the following closed form for all correlators of boxes in a line:
\beq\label{En}
\boxed{
E_n=\frac{1}{\mu}\big[ \big(\frac{2}{\mu-1}\big)^{n-1} - \big(\frac{-2}{\mu+1}\big)^{n-1} \big]
}\eeq
where $\mu=\sqrt{5+4\sqrt{2}}\approx 3.2645$. Notice that $E_n>0$ for all $n\geq2$. Obviously, flipping all outputs would change the sign of $E_n$ for all odd $n$'s. Table I lists the 12 first values of $E_n$, remarkably they are all of the form $n+m\sqrt{2}$, with $n$ and $m$ integers.

It remains to prove that these correlators guarantee $q_n(\vec x)\geq0$ for all $\vec x$. 

First, notice that by inserting the recusion formula (\ref{Enp1}) into (\ref{qnp1}) one gets, after some algebra:
\beq\label{qone}
q_n(\vec 1_n)=(\sqrt{2})^{n-1}
\eeq
where $\vec 1_n=(+1,+1,...,+1)$, with $n$ \ $+1$'s.

Second, using $q_n(x_1..x_{j-1}+-+x_{j+3}..x_n)=0$ for all $2\leq j\leq n-3$ one sees that the line of boxes can be split in parts each time the output string changes sign. Without loss of generality we may assume $x_1=+1$. Then, either all outputs are $+1$, i.e. $\vec x=\vec 1_n$ in which case $q_n(\vec 1_n)=\sqrt{2}^{n+1}>0$, or the first output $-1$ happen at position $j_0$, i.e. $x_{k}=+1$ for all $k<j_0$ and $x_{j_0}=-1$. In such a case, the box at position $j_0+1$ can be removed, as anyway the output $-1$ is certain. But if the $j_0+1$ box is removed, then the independence assumption implies that the probability factorizes: $p_n(x_1x_2...x_n)=p_{j_0}(++...++-)\cdot p_{n-j_0-1}(x_{j_0+2}...x_n)$. 

Eventually, all probabilities are products of the following terms: $p_n(\vec1_n)$, $p_n(\vec1_{n-1},-)$, $p_n(-,\vec1_{n-1})$ and $p_n(-,\vec1_{n-2},-)$. Remains to prove that these 4 terms are all non-negative for all $n$. Relation (\ref{qone}) proves the positivity of the first term and will be heavily used to prove the positivity of the 3 other term:
\beqa
p_{n+1}(\vec1_n,-)&=&p_{n+1}(-,\vec1_n)=p_n(\vec1_n)-p_{n+1}(\vec1_n,+) \nonumber\\
&=&\frac{1}{\sqrt{2}^{n+1}}\big(1-\frac{1}{\sqrt{2}}\big)>0 \\
p_{n+2}(-,\vec1_n,-)&=&p_n(\vec1_n)-p_{n+2}(-,\vec1_n,+) \nonumber\\
&-&p_{n+2}(+,\vec1_n,-)-p_{n+2}(+,\vec1_n,+)  \\
&=&\frac{1}{\sqrt{2}^{n+1}}\big(1-\frac{2}{\sqrt{2}}(1-\frac{1}{\sqrt{2}})-\half\big) \\
&>&0 \nonumber
\eeqa

In summary, all the correlators (\ref{En}) corresponding to $n$ boxes in a line are fixed by the assumptions $E_1=0$ and $E_2=\sqrt{2}-1$ (and $E_3\geq0$) and they guarantee that all probabilities $p(\vec x)\geq0$, for all output strings $\vec x$.
In the next section we prove that these correlators are also compatible with $n$ boxes in any polygon configuration.

\section{Polygons}\label{loop}
The smallest closed loop has a single box fed by the two links produced by a single source, see Fig. 1a. In this case we assume, similarly to a single box in a line, $E_1^o=0$, where the upper index $o$ indicates a closed loop. Accordingly, single boxes always produce fully random outputs. The second shortest loop has 2 boxes and 2 sources. Since $E_2^o$ is not limited by the NSI principle  (see Fig. 1), we assume it takes the largest possible value\footnote{At first sight, one may fear that combining the two end sources of the 2-box configuration in a line into a single source, thus changing the line into a loop configuration, leads to signaling, since $E_2\neq E_2^o$. However, the change in the source takes time to influence the boxes, the time of flight of the subsystems emitted by the sources. Hence, $E_2\neq E_2^o$ does not lead to signaling.}
: $E_2^o=1$.

Let's now consider polygons with $n+1$ vertices and edges, for $n\geq 2$. Using a similar technique as for the line we have:
\beqa\label{qo}
q^o_{n+1}(x_1..x_{n+1})&=& q_n(x_1..x_n) + \Pi_{j=1}^{n+1}x_j\cdot E^o_{n+1} \nonumber\\
&+& \sum_{k=1}^n E_k \cdot\sum_{\ell=1}^k \big(\Pi_{j=n+1-k+\ell}^{n+\ell}~ x_j\big) \nonumber\\
&\cdot& q_{n-1-k}(x_{n-1-k+\ell}..x_{n+\ell+2}) 
\eeqa
where all indices of $x_j$'s are assumed modulo $n+1$ (e.g. $x_{n+\ell+2}=x_{\ell+1}$) and we set $q_{-1}=q_0=1$.

Using $q_3(+-+)=0$ we deduce that for all $n\geq3$ $q_{n+1}(\vec 1_n,-)=0$. Let's apply this to the above formula:
\beqa
q^o_{n+1}(\vec 1_n,-)&=&q_n(\vec 1_n)-E^o_{n+1}  \nonumber\\
&+&\sum_{k=1}^n E_k\cdot\sum_{\ell=1}^k x_1..\hat x_\ell..\hat x_{\ell+n-k}..x_{n+1} \nonumber\\
&\cdot& q_{n-1-k}(x_{\ell}..x_{\ell+n-k-1}) \label{qno}\\
&=&\sqrt{2}^{n-1}-E^o_{n+1}-\sum_{k=1}^{n-2} E_k\cdot k\cdot\sqrt{2}^{n-k-2} \nonumber\\
&-&(n-1)E_{n-1} - nE_n=0
\label{qooo}
\eeqa
Consequently:
\beqa\label{Eo}
E^o_{n+1}&=&\sqrt{2}^{n-1}-\sum_{k=1}^{n-2}E_k\cdot k\cdot\sqrt{2}^{n-k-2} \nonumber\\
&-&n\cdot E_n - (n-1)\cdot E_{n-1} 
\eeqa
Using eq. (\ref{En}), with the same $\mu=\sqrt{5+4\sqrt{2}}$, one gets, for all $n\geq2$:
\beq\label{Eno}
\boxed{
E_n^o=\big(\frac{2}{\mu-1}\big)^n + \big(\frac{-2}{\mu+1}\big)^n
}\eeq
Note that $E_n^o>0$ for all $n$. Table I lists the 12 first values of $E_n^o$, remarkably they are all of the form $n+m\sqrt{2}$, with $n$ and $m$ integers, as we found for the correlators in a line.

Remains to prove that $q_{n+1}^o(\vec x)\geq 0$ for all $\vec x$. First, assume that not all outputs are equal. Then, there are 2 adjancent boxes with outputs $+1-1$. Since $p_3(+-+)=0$, we can remove the next box as it necessarily outputs $-1$. In this way one opens the loop and reduces it to a line for which we already proved non-negativity of all probabilities. Next, assume all outputs are +1. Since $E_n^o\geq0$ for all $n$, $p_n^o(\vec 1_n)\geq0$. Finally, assume all outputs are -1:
\beq
p_n^o(\overrightarrow{-1}_n)=\frac{1}{2^n}\big(1+n\sum_{k=2}^{n-1}(-1)^kE_k + (-1)^nE_n^o\big)
\eeq
From (\ref{En}) and (\ref{Eno}) straightforward computations shows that $E_{2n}-E_{2n+1}>0$ and $1+E_{n-1}-E_n^o>0$. Hence, $p_n^o(\overrightarrow{-1}_n)\geq0$ for all $n$.

\section{conclusion}
We proved that under the natural assumption $E_1=E_1^o=0$ there is a single pair of source and box in binary networks that maximizes the 2-box correlators $E_2$ and $E_2^o$, in the sense that the statistics produced by these resources in any binary network is uniquely defined. The existence and uniqueness of such a source and box is highly non-trivial. Admittedly, the deep reason why such a source and box exist and - especially - are unique is left open. Another question is whether this binary non-local resource can be realized within quantum theory and, if not, how close quantum can come?

A natural challenge is whether a similar result holds also for boxes with more outputs,  boxes connecting more than 2 sources, and/or sources connecting more than 2 boxes. We investigated numerically the case of boxes with 4 outputs under the natural assumption of output permutation symmetry, i.e. for all permutations $\pi$ of $\{1,2,3,4\}$, $p_n(\vec x)=p_n(\pi(\vec x))$ (same permutations for all outputs). However, we found no sign of the uniqueness for 4-output network boxes, though the question remains open and should be investigated in future work. Also, other types of networks should be studied, in particular configurations beyond lines and loops, like, e.g. star networks.

Network non-locality is a new and fascinating form of non-locality. It combines non-locality and ``entangling" joint measurements. It deserves to be analyzed within quantum theory and, as we do here, from outside quantum theory under the very general NSI principle: no-signaling and independence. The presented source and box for binary networks is a conceptual tool, similar in spirit to the PR- boxes introduced by Popescu and Rochlich \cite{PR}; such conceptual tool are useful to understand and simulate quantum correlations, and for applications in the spirit of device independent quantum information processing.

Among the many fascinating research directions to which this one contributes are all questions on the limits of quantum non-locality (e.g. information causality \cite{infoCausality}). In particular, macroscopic locality \cite{Miguel} and classical limits of non-local boxes \cite{Daniel,NGMagnets} are especially interesting and should be applid to network boxes.
\\

\onecolumngrid
\begin{table}[h]
	\centering
		\begin{tabular}
			{c|c|c|c|c|c|c|c|c|c|c|c|c}
			$n$ & 1 & 2 & 3 & 4 & 5 & 6 & 7 & 8 & 9 & 10 & 11 & 12  \\
			 \hline		
 $E_n$ & 0 & $\sqrt{2}-1$ & $3-2\sqrt{2}$ & $3\sqrt{2}-4$ & $3-2\sqrt{2}$ & $3-2\sqrt{2}$ & $10\sqrt{2}-14$ & $27-19\sqrt{2}$ & $22\sqrt{2}-31$ & $10-7\sqrt{2}$ & $51-36\sqrt{2}$ & $104\sqrt{2}-147$ \\
\hline
 $E_n^o$ & 0 & 1 & $2-\sqrt{2}$ & $4\sqrt{2}-5$ & $9-6\sqrt{2}$ & $6\sqrt{2}-8$ & $\sqrt{2}-1$ & $23-16\sqrt{2}$ & $37\sqrt{2}-52$ & $71-50\sqrt{2}$ & $32\sqrt{2}-45$ & $44-62\sqrt{2}$  \\
		\end{tabular}
\caption{\it Table of the first 12 correlators in line, $E_n$, and in a loop, $E_n^o$. (The indicated values of $E_2^o$ and $E_3^o$ are the maximal values compatible with $p_2^o(x_1x_2)\geq0$ and $p_3^o(x_1x_2x_3)\geq0$, respectively.)}
\end{table}

\twocolumngrid

\small
\section*{Acknowledgment} 
Stimulating discussions with Nicolas Brunner and financial support by the Swiss NCCR SwissMap are acknowledged.

\newpage
\appendix
\section{Proof of Lemma 1}
In this appendix we prove the lemma used in section III:

\vspace{0.3cm}\!\!\!\!\!\!\textbf{Lemma 1.}\quad \textit{In a line configuration with 7 boxes, whenever $E_1=0$ and $E_2=\sqrt{2}-1$, the following inequality holds:}
\begin{equation}
E_3^2 \geq (3-2\sqrt{2})^2.
\end{equation}

\begin{proof}
Let us identify 17 probabilities:
\begin{eqnarray}
P_1 &= P(-------)\\
P_2 &= P(---+---)\\
P_3 &= P(--+-+--)\\
P_4 &= P(+-+-+-+)\\
P_5 &= P(++-+-++)\\
P_6 &= P(+++-+++)\\
P_7 &= P(+++++++)\\
P_8 &= P(-++-+-+)\\
P_9 &= P(-+-+--+)\\
Q_1 &= P(----+-+)\\
Q_2 &= P(----+++)\\
Q_3 &= P(---+-++)\\
Q_4 &= P(---++++)\\
Q_5 &= P(--+-+++)\\
Q_6 &= P(--++-+-)\\
Q_7 &= P(-+-++++)\\
Q_8 &= P(+-+--++).
\end{eqnarray}
We define three vectors in $\mathbb{R}^{24}$:
\begin{equation}
\begin{split}
u = (8,9,6,1,9,4,5,3,7,1,6,2,7,1,7,5,3,7,1,6,2,6,7,1)\\
v = (4,2,7,6,4,2,2,4,1,7,6,8,6,3,5,4,2,7,1,1,3,5,3,5)
\end{split}
\end{equation}
and
\begin{equation}
\begin{split}
c = &\Big(\frac{11146 + 16545\sqrt{2}}{1160712},
\frac{2581 + 4686\sqrt{2}}{773808},
\frac{3 + 34\sqrt{2}}{18424},\\
&\frac{109 + 2003\sqrt{2}}{515872},
\frac{10253 + 16404\sqrt{2}}{1160712},
\frac{9529 + 14340\sqrt{2}}{1547616},\\
&\frac{29063 + 10799\sqrt{2}}{3095232},
\frac{50719 - 4773\sqrt{2}}{9285696},
\frac{1517 + 304\sqrt{2}}{147392},\\
&\frac{9(139 + 40\sqrt{2})}{147392},
\frac{3(-38 + 337\sqrt{2})}{257936},
\frac{6 + 19\sqrt{2}}{5488},\\
&\frac{227 + 1805\sqrt{2}}{515872},
\frac{16154 + 2677\sqrt{2}}{3095232},
\frac{107(80 + 139\sqrt{2})}{3095232},\\
&\frac{127277 - 41427\sqrt{2}}{9285696},
\frac{41879 + 4049\sqrt{2}}{3095232},
\frac{1387 + 366\sqrt{2}}{147392},\\
&\frac{1381 + 298\sqrt{2}}{147392},
\frac{3 + 34\sqrt{2}}{18424},
\frac{44 + 25\sqrt{2}}{5488},\\
&\frac{3(636 + 299\sqrt{2})}{257936},
\frac{2066 - 383\sqrt{2}}{442176},
\frac{8536 + 9995\sqrt{2}}{3095232}\Big)
\end{split}
\end{equation}
A lenghty but direct computation shows that
\begin{equation}
2^{14}\cdot\sum_{i=1}^{24} c_i P_{u_i} Q_{v_i} = 12\sqrt{2} - 17 + E_3^2.
\end{equation}
In other words, this expression does not involve any correlator with more than $3$ parties. Since the probabilities and all components of $c$ are positive, this expression also is positive, i.e.
\begin{equation}
12\sqrt{2} - 17 + E_3^2 \geq 0,
\end{equation}
which is equivalent to Eq.~\eqref{E3bound}.
\end{proof}

The proof above was derived by considering probabilities for seven parties in a line in the space of correlator monomials. According to~\eqref{qnp1} and~\eqref{En}, every probability in this scenario can be expressed as a linear combination of some products of correlators. The task of finding a linear combination of probabilities with positive coefficients for which all monomial except $E_3^2$ vanish then amounts to solving a linear program. The proof constitutes a solution of this program.

\end{document}